\documentstyle[preprint,eqsecnum,aps]{revtex}
\begin{document}
\draft
%
%
\title{
Two-Flavor Staggered Fermion Thermodynamics at $N_t = 12$
}
\author{Claude Bernard}
\address{
Department of Physics, Washington University, 
St.~Louis, MO 63130, USA
}
\author{Tom Blum}
\address{
Brookhaven National Laboratory, Upton, NY 11973-5000, USA
}
\author{Carleton DeTar}
\address{
Department of Physics, University of Utah, 
Salt Lake City, UT 84112, USA
}
\author{Steven Gottlieb and Kari Rummukainen}
\address{
Indiana University, Bloomington, IN 47405, USA
}
\author{Urs M.~Heller}
\address{
SCRI, The Florida State University, Tallahassee, FL 32306-4052, USA
}
\author{James Hetrick and Douglas Toussaint}
\address{
Department of Physics, University of Arizona, Tucson, AZ 85721, USA
}
\author{Robert L.~Sugar}
\address{
Department of Physics, University of California, 
Santa Barbara, CA 93106, USA
}
%
%
\date{\today}
\maketitle
\begin{abstract}
   We present results of an ongoing study of the nature of the high
temperature crossover in QCD with two light fermion flavors.  These
results are obtained with the conventional staggered fermion action at
the smallest lattice spacing to date---approximately 0.1 fm.  Of
particular interest are (1) a study of the temperature of the
crossover, an important indicator of continuum scaling, (2) a
determination of the induced baryon charge and baryon susceptibility,
used to study the dissolution of hadrons at the crossover, (3) the
scalar susceptibility, a signal for the appearance of soft modes, and
(4) the chiral order parameter, used to test models of critical
behavior associated with chiral symmetry restoration.  From our new
data and published results for $N_t = 4$, 6, and 8, we determine the
QCD magnetic equation of state from the chiral order parameter using
O(4) and mean field critical exponents and compare it with the
corresponding equation of state obtained from an O(4) spin model and
mean field theory.  We also present a scaling analysis of the Polyakov
loop, suggesting a temperature dependent ``constituent quark free
energy.''
\end{abstract}
\narrowtext
%
\section{Introduction}

Lattice simulations of high temperature QCD provide, at present, our
only firmly grounded theoretical insights into the phenomenology of
the transition from hadronic matter to the quark-gluon plasma, and
into the nature of the plasma itself.  Our ultimate goals in lattice
simulations at high temperature include (1) establishing continuum
scaling, important not only for determining the temperature of the
crossover, but necessary for the validity of all dynamical fermion
simulations, (2) establishing the mechanism for the
dissolution/formation of hadrons at the crossover, (3) exploring the
intricate critical behavior associated with the phase transition, and
(4) obtaining a quantitative characterization of the quark-gluon
plasma, including the equation of state.  To achieve these goals
requires a combination of advances in algorithms and computing power
\cite{ref:reviews,ref:reviews_detar}.  Recent improvements in quenched 
lattice algorithms hold promise for dynamical fermion simulations
\cite{ref:maclep,ref:improve}.  Here we present results of simulations 
with the conventional staggered fermion action at the smallest lattice
spacing to date.

The most detailed previous simulations with two-flavors of staggered
fermions were carried out at $N_t = 4$ ($a \approx 0.3$
fm)\cite{ref:kl}, $N_t = 6$ ($a \approx 0.2$ fm) and
\cite{ref:nt6} $N_t = 8$ ($a \approx 0.15$ fm) \cite{ref:nt8}.  
Results of those simulations reaffirmed the hypothesis that for two
flavors there is a rapid crossover, but no phase transition at nonzero
quark mass.  Moreover, the ratio $T_c/m_\rho$ was found to be
consistent with previous measurements.  Despite the lack of surprises
at $N_t = 6$ and 8, there are strong reasons to push to still smaller
lattice spacing.  The full flavor symmetry in the staggered fermion
scheme is restored only in the continuum limit.  Sigma models of
chiral symmetry restoration suggest that the crossover becomes a first
order phase transition as the number of flavors is increased
\cite{ref:chiral}.  If flavor symmetry breaking were to cause an
effective undercounting of quark flavors, one might expect a more
pronounced crossover, or even a genuine phase transition at smaller
lattice spacing.  Moreover, strong coupling distortions in the hadron
spectrum at $N_t = 6$ undermine credibility in a determination of the
crossover temperature or even in the plausibility of having achieved a
scaling ratio $T_c/m_\rho$.  For these reasons we undertook a
simulation at  $a \approx 0.1$ fm.  Preliminary results were
reported at the Bielefeld and Melbourne Lattice conferences
\cite{ref:reviews_detar,ref:sugar_lat94,ref:detar_lat95}.

The simulation with two flavors was carried out using the R algorithm
described in \cite{ref:rmd} at two quark masses, $am_q = 0.008$ and
0.016 and six couplings, $6/g^2 = 5.65$, 5.70, 5.725, 5.75, 5.80,
5.85, except that the 5.725 coupling was not simulated at the higher
mass. In each case the simulation was extended to at least 2000
molecular dynamics time units.  Lattices were saved at intervals of 8
time units.  For the present analysis the first 500 time units were
omitted for equilibration.  Most of the results reported here are
based on an analysis of the approximately 180 remaining lattices at
each parameter pair.  For the sake of comparison, we also present some
new results for $N_t = 6$, based on a compilation of lattices from our
equation of state study \cite{ref:eos6}.

In addition to the standard variables, the chiral order parameter and
the Polyakov loop, we introduce a fuzzy Polyakov loop in an effort to
regulate the ultraviolet divergence that becomes increasingly
troublesome at small lattice spacing.  We also measure the baryon
susceptibility, the induced quark number, and the disconnected chiral
susceptibility.  Results are presented in Sec.~2.  We show that with
our choice of masses and couplings, the crossover is more clearly
defined in terms of the baryon susceptibility and induced quark
number.  In Sec.~3 we present a scaling analysis of the Polyakov loop
variable for world data ranging from $N_t = 4$ to $N_t = 12$, in terms
of a constituent quark free energy.  We also present a critical
scaling analysis of the dependence of
$\left\langle\bar\psi\psi\right\rangle$ on temperature and quark mass
(magnetic equation of state), including a compilation of world data
for this variable.  This analysis is carried out in the context of
mean field as well as $O(4)$ critical behavior.

\section{Locating the Crossover}

\subsection{Polyakov loop, $\left\langle\bar\psi\psi\right\rangle$, 
fuzzy loop}

The high temperature crossover is conventionally located from the
inflection point in a plot of the Polyakov loop or chiral order
parameter as a function of coupling.  Figures \ref{fig:pbpvsbeta} and
\ref{fig:repvsbeta} plot these quantities at fixed bare quark mass as
a function of coupling.  It is clear that despite the small errors, at
these quark masses the effect of the the crossover on these quantities
is subtle, indeed.  

The Polyakov loop measures the change $f(T,m_q)$ in the ensemble
free energy caused by the introduction of a static test quark.  In
particular, with our normalization
\begin{equation}
   \left\langle\mathop{\rm Re} P\right\rangle = 3 \exp[-f(T,m_q)/T]
\end{equation}
As the lattice spacing shrinks, the test quark self-energy develops an
ultraviolet divergence, which may overwhelm a crossover signal in this
quantity.  Thus one might hope that by increasing the radius of the
test charge, one might regulate the divergence and recover a signal at
least as strong as was seen on a coarser lattice.  Accordingly, we
constructed a ``fuzzy Polyakov loop'' variable in analogy with
techniques introduced for glueball sources \cite{ref:fuzz}.
To do so, we replaced the conventional product of forward links
\begin{equation}
\left\langle\mathop{\rm Re} P({\bf x})\right\rangle = 
  \left\langle\mathop{\rm Tr} 
    \prod_{t=0}^{N_t - 1} U_t({\bf x}, t)\right\rangle
\end{equation}
with
\begin{equation}
\left\langle\mathop{\rm Re} F({\bf x})\right\rangle = 
\left\langle\mathop{\rm Tr}
   \prod_{t=0}^{N_t - 1} F_t({\bf x}, t)\right\rangle
\end{equation}
where
\begin{equation}
  F_t({\bf x},t)  = \alpha U_t + \beta \sum U_{\rm staple}
\end{equation}
The six staples associated with the link $U_t({\bf x},t)$ are the
usual three-link products of the form
\begin{equation}
  U_x({\bf x},t)U_t({\bf x} + \hat x,t)U^\dagger_x({\bf x},t + 1)
\end{equation}
centered on the link from $({\bf x},t)$ to $({\bf x},t+1)$.  A
weighting $\alpha = \beta = 1/7$ was determined from a rough
optimization of the variance.  By construction the weights sum to 1
and we do not project the matrix $F_t$ onto SU(3), so this observable
still creates a source of precisely one color triplet. 

Figure \ref{fig:refvsbeta} gives the result.  Comparing with the
standard Polyakov loop in Fig.~\ref{fig:repvsbeta}, we see that
smearing indeed reduces the relative statistical error, but apparently
does not enhance the crossover signal.

\subsection{Baryon susceptibility}

The conventional Feynman path integral simulates the grand canonical
ensemble in baryon number at zero chemical potential.  The baryon
susceptibility measures fluctuations in the baryon number of the
ensemble.  It is defined as the derivative of the baryon charge
density with respect to chemical potential.  The susceptibility can be
defined separately for each flavor.  Thus with two quark flavors, two
susceptibilities can be measured: a flavor singlet and flavor
nonsinglet \cite{ref:barsus}.  Both quantities are obtained at zero
chemical potential.
\begin{equation}
   \chi_{\rm s,ns} =  \left(\frac{\partial}{\partial \mu_u} \pm
                 \frac{\partial}{\partial \mu_d}\right)
                  (\rho_u \pm \rho_d)
\end{equation}
The nonsinglet susceptibility is compared with results for lower $N_t$
\cite{ref:barsus} in Fig.~\ref{fig:xnsvsbeta}.  
Also indicated is the free lattice quark value for each $N_t$.  A
common feature is the abrupt rise in susceptibility at crossover,
followed by an asymptotic approach to the free lattice quark value.
At lower values of $N_t$, where the crossover has been located with
traditional methods, we find that the baryon susceptibility reaches
1/3--1/2 of the free quark asymptotic value at the crossover.  Since
this observable is based on a conserved charge, it is not
renormalized, and offers our most distinctive signal for the
crossover.  Following the same rule for $N_t = 12$ places the
crossover in the $am_q = 0.008$ series between $6/g^2 = 5.65$ and 5.70
and in the $am_q = 0.016$ series between $6/g^2 = 5.75$ and 5.80.

To convert these results to a temperature, we use a scale in which the
rho meson mass is taken to be 770 MeV, regardless of quark mass.  The
rho mass in lattice units is obtained in turn from an empirical fit to
masses in a compilation of two-flavor staggered fermion spectral
simulations \cite{ref:eos4}, and includes an extrapolation beyond the
parameter range of spectral measurements ($6/g^2 > 5.7$) using
tadpole-improved asymptotic scaling.  Details are given in the
Appendix.  From the baryon susceptibility we then place the crossover
at $T_c = 143-154$ MeV$_\rho$ at the lighter quark mass and $T_c =
142-150$ MeV$_\rho$ at the heavier quark mass.
\subsection{Induced quark number}

Another confinement-sensitive observable is the induced quark number
\cite{ref:barcorr}.  This observable measures the total residual
light-quark number in an ensemble containing a single test quark.
\begin{equation}
   Q_{\rm ind} = \int \rho_{\rm ind}(r) d^3r
\end{equation}
The induced quark number density $\rho_{\rm ind}$ is measured in the
presence of the test quark.  Operationally, the quantity measures the
correlation between the Polyakov loop and the light-quark density
\cite{ref:barcorr}.  

This observable is subject to considerable fluctuation.  We found it
particularly effective to introduce the test charge through the fuzzy
Polyakov loop variable described above.  The baryon density of the
dynamical quark is computed using a random source estimator
\cite{ref:barcorr}.  To improve the signal further, we adjusted the
number of random sources using an adaptive procedure on a
configuration-by-configuration basis as follows: starting with a
minimum of 20 random sources the variance of the total induced baryon
number was estimated.  If the variance was greater than tolerance,
another 20 random sources were added to the sample, and so on, up to a
maximum of 80 for one configuration.

Results are shown in Fig.~\ref{fig:qvsT} and compared with results
from simulations at lower $N_t$.  The induced quark number is expected
to be exactly $-1$ at zero temperature, since confinement requires
screening of the test charge by a single antiquark.  At the crossover,
this quantity rises rapidly, approaching zero in the high temperature
phase.  At lower $N_t$ the induced quark number reaches approximately
$-0.1$ at crossover.  Thus the induced quark number density can give
an operational definition of the crossover.  Applying this rule to the
$N_t = 12$ data gives $6/g^2 = 5.65-5.70$ ($T = 143-154$ MeV$_\rho$)
in the $am_q = 0.008$ series and $6/g^2 = 5.70-5.80$ ($T = 134-150$
MeV$_\rho$) in the $am_q = 0.016$ series.  This crossover location is
consistent with, but somewhat less precise than, that found from the
baryon susceptibility.
%
%
\subsection{Chiral Susceptibility}

Another signal for the crossover is the singlet chiral susceptibility
\begin{equation}
   \chi_m = \left.\frac{\partial \left\langle\bar\psi\psi\right\rangle}
     {\partial m}\right|_{6/g^2}
\end{equation}
which measures fluctuations in the chiral order parameter
$\left\langle\bar\psi\psi\right\rangle$
\cite{ref:kl}.  Here we differentiate with respect to equated up and
down quark masses.  Since the derivative is the space-time integral of
the correlator $\left\langle \bar \psi \psi(r) \bar \psi
\psi(0)\right\rangle$, a peak in this observable occurs at a minimum
in the $\sigma$ meson screening mass, indicating the presence of a
soft mode.  Such soft modes are expected in models of critical
behavior \cite{ref:chiral}. Like the $\sigma$ meson propagator, the
singlet chiral susceptibility can be decomposed into two
contributions: quark-line-connected and quark-line-disconnected:
\begin{equation}
  \chi_m = \chi_{\rm conn} + \chi_{\rm disc}.
\end{equation}
The connected susceptibility in another guise is the derivative of
$\left\langle\bar\psi\psi\right\rangle$ with respect to valence quark
mass at fixed sea quark mass and $6/g^2$ \cite{ref:CC}, while the
disconnected susceptibility is the configuration variance of
$\left\langle\bar\psi\psi\right\rangle$:
\begin{equation}
  \chi_{\rm disc} = 
  \left\langle \left\langle\bar\psi\psi\right\rangle_{\rm conf}^2 
     \right\rangle_U - 
  \left\langle \left\langle\bar\psi\psi\right\rangle_{\rm conf} 
     \right\rangle_U^2
\label{eq:xsdisc}
\end{equation}
where $\left\langle\bar\psi\psi\right\rangle_{\rm conf}$ is the
expectation value computed on a single gauge configuration and
$\left\langle \right\rangle_U$ denotes averaging over all gauge
configurations in the sample.

We used the standard Gaussian random source estimator for
$\left\langle\bar\psi\psi\right\rangle_{\rm conf}$,
\begin{equation}
  \left\langle\bar\psi\psi\right\rangle_{\rm conf} = 
     \langle \xi^\dagger S^{-1} \xi \rangle_\xi
\end{equation}
One must take care in using this estimator for Eq.~(\ref{eq:xsdisc}),
since the square involves correlated quartic terms in the random
source variable.  Averaging then reintroduces part of the connected
susceptibility along with the disconnected susceptibility.  We chose
to avoid the problem altogether by computing the estimator from five
separate random sources on each configuration and forming products in
Eq.~(\ref{eq:xsdisc}) only from pairs of estimators with different
random sources.  (For the smaller volume $N_t = 6$ data we used 33
random sources per configuration.)

Results for $\chi_{\rm disc}$ and $\chi_{\rm conn}$ are compared in
Figs.~\ref{fig:xsdiscvsT} and \ref{fig:xsconnvsT} with the results of
Karsch and Laermann at $N_t = 4$ \cite{ref:kl} and results from a
reanalysis of a set of $N_t = 6$ lattices retained from an equation of
state study\cite{ref:eos6}.  The benefits of the larger sample in the
$N_t = 4$ data are clearly evident in this variable.  At the lighter
quark mass ($am_q/T = 0.08$) the $N_t = 4$ octagons define a crossover
peak at about 145 MeV$_\rho$. This peak can be sought in the $N_t = 6$
data at a comparable quark mass ($am_q/T = 0.075$).  The data are
crude but not inconsistent with a comparable peak.  The bursts plot
$N_t = 12$ data at the lighter quark mass ($am_q/T = 0.096$), which
are not inconsistent with a peak in the range 140-160 MeV$_\rho$.  The
signal in the higher mass $N_t = 12$ data (diamonds) is inconclusive.

That the signal in $\chi_{\rm disc}$ should be weaker at higher quark
mass is expected from models of critical behavior that place the
critical point at zero quark mass.  The mass ratio $m_\pi^2/m_\rho^2$
measures proximity to the critical point.  In the $N_t = 4$ data, this
ratio is less than 0.1 for the lighter quark mass ($am_q = 0.020$)
data and between 0.1 and 0.2 for the $am_q = 0.0375$ data.  For the
$N_t = 6$ data it is also between 0.1 and 0.2.  For the $N_t = 12$
data, it is between 0.2 and 0.3 for $am_q = 0.008$ and 0.3 to 0.4 for
$am_q = 0.016$.  Thus we expect a weakening of the crossover signal at
the quark masses treated in our higher $N_t$ data.

The connected part of the susceptibility $\chi_{\rm conn}$ is compared
in Fig.~\ref{fig:xsconnvsT} with results from $N_t = 4$ and our
results from $N_t = 6$.  This quantity is far less noisy than the
disconnected contribution.  The lack of a discernible peak in the $N_t
= 12$ results is presumably another consequence of the greater distance
from the critical point.
\subsection{Crossover location}

To summarize, the observables we have considered place the crossover
at $N_t = 12$ in the range $6/g^2 = 5.65-5.70$ at $am_q = 0.008$ and
in the range $5.75-5.80$ at $am_q = 0.016$, which we translate to
$143-154$ MeV$_\rho$ and $142-150$ MeV$_\rho$, respectively.  This
determination is compared with results for a wide variety of
two-flavor simulations with both Wilson and staggered fermions in
Fig~\ref{fig:tcovermrhopi2sq} \cite{ref:sugar_lat94,ref:tcmrho}.  Here
we plot the temperature in units of the $\rho$ meson mass {\it vs} the
square of the ratio of the $\pi$ to $\rho$ mass.  For the staggered
fermion simulations, we use the lightest local (non-Goldstone) pion.
This state is expected to be degenerate with the Goldstone pion in the
continuum limit, so progress toward the origin in this ratio measures
restoration of flavor as well as chiral symmetry.  Except for a few
$N_t = 4$ points from Wilson fermions, the consistency of these
results (within about ten percent) is quite striking.

\section{Scaling Tests}

\subsection{Constituent Quark Free Energy}

The wide range of $N_t$ values now available makes possible an
amusing analysis of the Polyakov loop variable, which measures the
change in the free energy of the thermal ensemble due to the
introduction of a point spinless test quark.  This free energy
difference,
\begin{equation}
    f(T,m_q) = -T\log\langle\mathop{\rm Re} P/3\rangle, 
\end{equation} 
a function of the temperature $T$ and light quark mass $m_q$, includes
the lattice-regulated ultraviolet-divergent self-energy of the point
source, proportional to the inverse lattice spacing $1/a$, and the
free energy of the screening cloud of light antiquarks, quarks, and
gluons, which we might call the ``constituent quark free energy''.
Computing the self energy to leading order in perturbation theory, we
have
\begin{equation}
    f(T,m_q) = 2\pi C_F\alpha_V\gamma/a + f_{\rm cq}(T,m_q).
\label{eq:cqfreeen}
\end{equation}
where $C_F = 4/3$ is the color Casimir factor for the triplet
representation, $\alpha_V$ is the color fine structure constant
appropriate for heavy quark bound states at the same lattice scale,
and
\begin{equation}
  \gamma = \frac{1}{N_s^3} \sum_k \frac{1}
     {6 - 2\sum_{\mu=1}^3 \cos(2 \pi k_\mu/N_s) + (Ma)^2} 
\end{equation}
is the dimensionless static lattice propagator for a Debye-screened
electrostatic gluon field evaluated at zero separation.  Screening
provides an arbitrary infrared cutoff.  Although the ultraviolet
divergent contribution $1/a$ is uniquely determined in the limit $a
\rightarrow 0$, an infrared cutoff, here embodied in the Debye mass
$M$, determines where the contribution from the point quark ends and
the contribution from the screening cloud begins.  Thus this approach
offers no unique definition of a ``constituent quark free energy''.
Instead, within the framework of any consistent definition,
Eq.~(\ref{eq:cqfreeen}) permits a separation of two contributions, one
varying in a known way with the lattice scale $1/a$ and the other,
unknown, but scale-invariant (at least in the continuum limit).
Herein lies its predictive power: no matter what choice is made for
the infrared cutoff, a measurement of $\mathop{\rm Re} P$ at two
different values of $N_t$ can be used to predict $\mathop{\rm Re} P$
at any other $N_t$.

For present purposes, using $N_t = 1/aT$, we choose a simpler
approximate form
\begin{equation}
  f_{\rm cq}(T,m_q) = 
   -T\left(\log\langle\mathop{\rm Re} P/3\rangle + cN_t\right)
\end{equation} 
and adjust the dimensionless constant $c$ by eye to achieve the rough
scaling agreement shown in Fig.~\ref{fig:cqfen}.  For each data set
only values for the lightest available quark mass are used.  Although
the quark mass values $m_q/T$ are not the same from one $N_t$ to the
next in this figure, they are small ($m_q/T \le 0.1$) and the small
variation would be expected to make little difference (of the order 10
MeV) in the free energy.  The best value for $c$ appears to be about
0.4, compared with values ranging from 0.26 to 0.35 expected from the
lowest order perturbative self energy with a screening mass $M =
3.2T$.  It is curious that at the crossover, the free energy drops by
about the 300 MeV expected in a constituent quark model with
deconfinement at high temperature.

\subsection{Critical Behavior}

An SU(N)$\times$SU(N) sigma model has often been proposed as the
paradigm for the high temperature phase transition in QCD
\cite{ref:chiral}. For two light flavors ($N = 2$) one appealing
alternative, consistent with numerical results, is that chiral
symmetry restoration proceeds through a second order phase transition
at zero quark mass with attendant O(4) critical behavior.  Flavor
symmetry breaking in the staggered fermion scheme preserves only an
exact O(2) subgroup, so one may expect O(2) critical behavior on
coarse lattices.  Away from the critical point, where fluctuations are
unimportant, spin systems often exhibit mean field behavior.  Ko\'cic
and Kogut have even proposed that mean field behavior could extend up
to the critical point, as a consequence of the composite nature of the
scalar sigma model fields as they appear in QCD\cite{ref:kk}.  While
it remains to be established whether their arguments apply to QCD, we
consider the mean field alternative a plausible first approximation,
since we do not know the extent of the Ginzburg scaling region.  Thus
even if O(4) critical behavior is ultimately found in QCD at
sufficiently large volume, it is possible that results imitate mean
field behavior for lattices of the size considered in this study and
at temperatures and quark masses that are not sufficiently close to
the critical point.

Karsch \cite{ref:karsch} and Karsch and Laermann \cite{ref:kl} have
analyzed the crossover, starting from the Fisher scaling hypothesis
for the scaling of the critical contribution to the free energy
\begin{equation}
   f_{\rm crit}(t,h) = b^{-d}f_{\rm crit}(b^{y_t}t,b^{y_h}h).
\end{equation}
{}From this scaling behavior one derives a scaling relation for the
critical contribution to the magnetization $s = -\partial f_{\rm
crit}/\partial h$:
\begin{equation}
  s(t,h) = h^{1/\delta} y(x) 
\label{eq:eos}
\end{equation} 
where $x = th^{-1/\beta\delta}$ and $y(x)$ is a scaling function.  In
QCD the quark mass plays the role of the magnetic field and
$\left\langle\bar\psi\psi\right\rangle$, the magnetization.
Specifically, Karsch suggests using $h$ = $m_q/T = am_qN_t$ and $t =
6/g^2 - 6/g_c^2(0,N_t)$, where $g_c(0,N_t)$ is the critical gauge
coupling at zero quark mass for a particular $N_t$\cite{ref:karsch}.
This identification leads to a successful accounting for the crossover
location (``pseudocritical point'') $6/g_{\rm pc}(m_q)^2$ for $N_t =
4$ at small quark mass.

We extend the Karsch and Laermann analysis to a wider range of lattice
spacings and test the scaling relation (\ref{eq:eos}) directly.  To
anticipate the need to explore critical behavior, while simultaneously
approaching the continuum limit, we propose an identification that
avoids quantities with anomalous dimensions, but entails a
translation to physical units:
\begin{eqnarray}
   h &=& m_\pi^2(m_q,T=0)/m_\rho^2(m_q,T=0) \\
   t &=& [T - T_c(0)]/T_c(0) \\
   s &=& h^{-1}m_q\langle\bar\psi \psi(m_q,T)\rangle/T^4
\end{eqnarray} 
The scaling relation (\ref{eq:eos}) then gives a universal function
\begin{equation}
 y(x) =h^{-1-1/\delta}m_q\langle\bar\psi \psi(m_q,T)\rangle/T^4
\end{equation}
with $x = th^{-1/\beta\delta}$.  The extra factor $h^{-1}$ is needed
to compensate for the quark mass factor $m_q$.  Let us apply this
analysis to data for $\left\langle\bar\psi\psi\right\rangle$ from
several groups
\cite{ref:KS2thermo}.  We begin by applying the following cuts to the
world data:
\begin{itemize}
\item $m_q/T = am_qN_t < 0.2$
\item $h \le 0.5$ 
\item $-1 \le x \le 1$
\item $-0.5 \le t \le 0.5$
\end{itemize}
The first cut keeps parameters within range of the spectrum
interpolation.  The last three are an attempt to keep values within
the region dominated by the scaling part of the free energy.  Since we
have no a priori knowledge of the extent of this region, these choices
are arbitrary.  Indeed the cut on $h$ is probably too
generous, as we shall see, but we have set it high initially so that
all of the $N_t = 12$ data are included.

We first carry out a mean field analysis of the data.  The mean field
magnetic equation of state $y(x)$ is determined from the global
minimum of the usual quartic free energy
\begin{equation}
   F(y,x) = by^4 + axy^2 - y
\end{equation}
where $a$ and $b$ are adjustable scale parameters.  The third
adjustable parameter is the critical temperature $T_c(0)$ at zero
quark mass, which we vary in 10 MeV increments.  Our exploratory
strategy is to seek the best agreement between world data (after the
cuts listed above) and the resulting magnetic equation of state
$y(x)$, giving higher priority to data with higher $N_t$.  Setting the
critical exponents to their mean field values, $\delta = 1/3$ and
$\beta = 1/2$, and adjusting $T_c(0)$ to get the best qualitative
agreement according to these preferences, we get the result shown in
Fig.~\ref{fig:pbpMF140} with the choice $T_c = 140$.  Since the $N_t =
4$ data appear to be outliers, it was necessary to omit them from the
fit to the mean field magnetic equation of state.  Apart from the $N_t
= 12$ data at higher quark mass, the agreement is surprisingly good.

Our higher mass $N_t = 12$ data are strikingly inconsistent with
scaling.  Indeed, as we have remarked in our analysis of the
disconnected susceptibility, the $am_q = 0.016$ data are taken at
values of $h$ higher than those of the rest of the sample included in
the graph.  Thus it is plausible that these data are taken too far
from the critical region.

An alternative strategy, the one used in \cite{ref:reviews_detar}, is
to emphasize agreement among world data for the full range of $N_t$
without regard to the mean field magnetic equation of state.  Our best
result is shown in Fig.~\ref{fig:pbpMF150} with the choice $T_c =
150$.  The mean field curve, based on fitting only over the range
$-0.25 < x < 0.25$ clearly does not match the data as well as it did
with $T_c = 140$.

The corresponding exercise with O(4) critical exponents $\delta =
4.82$ and $\beta = 0.38$ is shown in Figs.~\ref{fig:pbpO4130} and
\ref{fig:pbpO4140}.  In this case the magnetic equation of state is not
known in closed form.  Instead, it was obtained from a simulation of
an O(4) spin model \cite{ref:O4spin}, using critical exponents and the
critical coupling reported by Kanaya and Kaya \cite{ref:KanayaKaya}.
The level of agreement is comparable to that of the mean field
analysis.

At the present level of precision it is not possible to distinguish
O(4) from mean field critical behavior, let alone from O(2) (not
shown).  Obviously a host of systematic errors, including finite
volume effects and deviations from continuum scaling enter the
analysis, so refinements are certainly needed before the method can
serve as a definitive test of critical behavior.


\section{Summary and Conclusions}

We presented results of a numerical simulation of QCD with
conventional staggered fermions at nonzero temperature at the smallest
lattice spacing attempted to date---approximately 0.1 fm.  Numerical
results are collected in Tables
\ref{tab:nt12m008a}---\ref{tab:nt12m016b}. Most of our data is
obtained in simulations at $N_t = 12$, but some new results are also
presented for $N_t = 6$.  Locating the high temperature crossover at
0.1 fm for $h = m_\pi^2/m_\rho^2 > 0.2$ is difficult in the
conventional observables $\langle \mathop{\rm Re} P \rangle$ and
$\left\langle\bar\psi\psi\right\rangle$.  However, the baryon
susceptibility and induced quark number show an abrupt rise consistent
with what is found at lower $N_t$.  We use these signals to locate the
crossover.  At the lighter quark mass it is found to be in the range
$143-154$ MeV$_\rho$, consistent with results at lower $N_t$.  Judged
in terms of the ratio $h$, our results at higher quark mass $am_q =
0.016$ are farther from the critical point than the world data at
smaller $N_t$.  Thus the expected crossover peak in the disconnected
chiral susceptibility is barely discernible in our $am_q = 0.016$ data
set, but it is present in our $am_q = 0.008$ data.  We demonstrated a
rough scaling analysis of the Polyakov loop variable in terms of a
constituent quark free energy.  Finally,
$\left\langle\bar\psi\psi\right\rangle$ shows remarkable agreement
among world data with appropriate cuts over the range $N_t = 6$, 8,
and 12 with a mean field magnetic equation of state and an assumed
crossover temperature of approximately 140 MeV {\em at zero quark
mass}.  Our analysis is not refined enough to distinguish among O(2),
O(4), and mean field scaling, however.

\section*{Acknowledgements}

We are grateful to Frithjof Karsch and Edwin Laermann for providing us
with tabulations of results from their $N_t = 4$ study.  We also thank
Edward Shuryak and Krishna Rajagopal for stimulating discussions.
This work was supported in part by the U.S.~National Science
Foundation under grants PHY91-16964 and PHY93-09458 and by the
U.S.~Department of Energy under contracts DE-FG02-91ER-40661,
DE2-FG02-91ER-40628, DE-FG03-95ER-40906, DE-FG05-85ER250000, and
DE-FG05-92ER-40742, and was carried out through grants of computer
time from the NSF at NCSA, SDSC, PSC, and by the DOE at NERSC and
Sandia.

\appendix
\section{Conversion from Lattice to Physical Units}

The lattice scale is converted to physical units using $a = a
m_\rho/770$ MeV$^{-1}$ following a method similar to that of
Ref.~\cite{ref:eos4}.  The numerator $a m_\rho$ is found through a
combination of interpolation and extrapolation from zero temperature
spectral measurements described below.  Of course, taking the physical
value to be 770 MeV ignores shifts in this value as the quark mass is
varied.  We also require masses of the Goldstone pion $m_\pi$ and the
second local pion $m_{\pi 2}$, which becomes degenerate with the pion
in the continuum limit.  Unfortunately, the highest value of $6/g^2$
for which spectral data are available is 5.7, but our $N_t = 12$ data
span the range 5.65 to 5.85, requiring extrapolation.  For data at
lower $N_t$ an interpolation suffices.  Figure \ref{fig:coverage}
indicates the parameter range of spectral data used for interpolation
\cite{ref:spectrum}.  In selecting the world data sample, we have not 
attempted to correct for finite volume effects.

For the interpolation we found a polynomial/spline expansion
convenient:
\begin{eqnarray}
  m_\pi/m_\rho = R_{\pi/\rho}(6/g^2,am_q) &=& 
                (am_q)^{1/2}S_{\pi}(6/g^2) 
                + am_q T_{\pi}(6/g^2) 
                + (am_q)^{3/2}U_{\pi}(6/g^2) \\
  a m_\rho = R_\rho(6/g^2,am_q) &=& S_{\rho}(6/g^2) 
                + am_q T_{\rho}(6/g^2) 
                + (am_q)^2 U_{\rho}(6/g^2) \\
  a m_{\pi 2} = R_{\pi 2}(6/g^2,am_q) &=& S_{\pi 2}(6/g^2) 
                + am_q T_{\pi 2}(6/g^2) 
                + (am_q)^2 U_{\pi 2}(6/g^2) \\
\end{eqnarray}
where the $S$, $T$, and $U$'s are natural cubic splines with three
knots, $6/g^2 = 5.3, 5.5, 5.7$ selected to span the range of available
spectral data.

For the $N_t = 12$ data over the range $6/g^2 = [5.65, 5.85]$, we
chose an extrapolation based on two-loop asymptotic scaling:
\begin{equation}
   a\Lambda(6/g^2) = 
   [8\pi^2/(3b_0 g^2)]^{b_1/b_0^2}\exp([-8\pi^2/(b_0 g^2)]
\end{equation}
where $b_0 = 11 - 2 N_f/3$ and $b_1 = 51 - 19 N_f/3$ and $N_f = 2$.
In place of the bare lattice coupling, we used the tadpole improved
value proposed by Mackenzie and Lepage \cite{ref:maclep} and
generalized to two flavors by Bitar et al.\ \cite{ref:beff_ks}:
\begin{equation}
  6/g^2_{\rm eff} = \beta_{\rm eff}(\Box) = -2/\log(\Box/3) - 1.99/\pi
\end{equation}
where $\Box$ is the plaquette variable (normalized to 3 for the
vacuum).  Applying the extrapolation requires knowledge of the
plaquette over the range of extrapolation.  Since zero temperature
values are unavailable over this range, we used the plaquette values
measured in our high temperature study, using a spline interpolation
formula in the same pattern as the masses above, but with knots at
$6/g^2 = 5.65, 5.75, 5.85$.  At our small lattice spacing, the
temperature effect on the plaquette is negligible for this purpose.
To avoid introducing distortions in our predicted mass values
resulting from a switch in method from the spline interpolation to the
extrapolation procedure, we elected to use only an extrapolation,
based on interpolated spectral data at $6/g^2 = 5.65$, the lowest
value in our data set, rather than at $6/g^2 = 5.70$, the highest
parameter for which spectral data is available.  One must allow for a
rescaling of the quark mass in the extrapolation.  Thus to determine
the starting point for extrapolation, we first solve for the scale
parameter $s$
\begin{equation}
   s = \frac{\Lambda\{\beta_{\rm eff}[\Box(6/g^2,am_q)]\}}
            {\Lambda\{\beta_{\rm eff}[\Box(5.65,am_q/s)]\}}.
\end{equation}
The extrapolated masses are then obtained by rescaling from the
interpolation at $6/g^2 = 5.65$:
\begin{eqnarray}
   m_\pi/m_\rho &=& R_{\pi/\rho}(5.65,am_q/s) \\
   a m_\rho &=& s R_\rho(5.65,am_q/s) \\
   a m_{\pi 2} &=& s R_{\pi 2}(5.65,am_q/s).
\end{eqnarray}
The measured mass values at 5.70 can then be used as a check of the
extrapolation.  In Fig.~\ref{fig:spectrum} we compare the empirical
formula for hadron masses with the measured values.  The extrapolation
from $6/g^2 = 5.65$ predicts slightly higher mass values at $6/g^2 =
5.7$ than measured. The agreement is certainly adequate for present
purposes.  To refine the method would require a more detailed study of
finite volume effects \cite{ref:finiteV}.





%
\begin{figure}
\caption{Chiral condensate $\left\langle\bar\psi\psi\right\rangle$ 
{\it vs} $6/g^2$.}%
\label{fig:pbpvsbeta} 
\end{figure}
\begin{figure}
\caption{Polyakov loop {\it vs} $6/g^2$.}
\label{fig:repvsbeta} 
\end{figure}
\begin{figure}
\caption{Fuzzy Polyakov loop {\it vs} $6/g^2$.}
\label{fig:refvsbeta} 
\end{figure}
\begin{figure}
\caption{Nonsinglet baryon susceptibility {\it vs} $6/g^2$ for $N_t =
4,6,8,12$.  Free lattice quark values for each $N_t$ are indicated by
horizontal lines on the right of the plot.}
\label{fig:xnsvsbeta} 
\end{figure}
\begin{figure}
\caption{Induced quark number. Octagons are for fixed lattice spacing.}
\label{fig:qvsT} 
\end{figure}
\begin{figure}
\caption{Disconnected chiral susceptibility. Crosses for $N_t =
4$, $am_q = 0.0375$\protect\cite{ref:kl}, octagons for $N_t = 4$,
$am_q = 0.02$\protect\cite{ref:kl}, squares for $N_t = 6$, $am_q =
0.0125$, diamonds for $N_t = 12$, $am_q = 0.016$, and bursts for $N_t
= 12$, $am_q = 0.008$.}
\label{fig:xsdiscvsT} 
\end{figure}
\begin{figure}
\caption{Connected chiral susceptibility.  Same notation as 
Fig.~\protect\ref{fig:xsdiscvsT}.}
\label{fig:xsconnvsT} 
\end{figure}
\begin{figure}
\caption{Crossover temperature in units of the rho meson mass {\it vs} the
square of the pi to rho mass ratio.  For staggered fermions the mass
of the lightest non-Goldstone pion is used \protect\cite{ref:tcmrho}.
The vertical line locates the physical ratio.}
\label{fig:tcovermrhopi2sq} 
\end{figure}
\begin{figure}
\caption{Constituent quark free energy as defined in text.}
\label{fig:cqfen} 
\end{figure}
\begin{figure}
\caption{Scaled $\langle \bar\psi \psi \rangle$ {\it vs} scaled temperature
with mean field critical exponents at $T_c(0) = 140$ MeV.}
\label{fig:pbpMF140}  
\end{figure}
\begin{figure}
\caption{Same but with $T_c(0) = 150$ MeV.}
\label{fig:pbpMF150} 
\end{figure}
\begin{figure}
\caption{Same but with O(4) critical exponents and $T_c(0) = 130$ MeV.}
\label{fig:pbpO4130} 
\end{figure}
\begin{figure}
\caption{Same but with O(4) critical exponents and $T_c(0) = 140$ MeV.}
\label{fig:pbpO4140} 
\end{figure}
\begin{figure}
\caption{Range of parameter pairs $(6/g^2, am_q)$ for two-flavor 
staggered fermion thermodynamic simulations (various plot symbols as
indicated) and spectral data (symbol ``S'') used to set the lattice
scale.}  
\label{fig:coverage} 
\end{figure}
\begin{figure}
\caption{Empirical fit to the $\pi$, $\rho$, and $\pi_2$ masses in the 
compilation of available spectral data. Open symbols (square, octagon,
diamond) and error bars show the respective measured values.  The
cross, burst, and plus show the respective empirical estimates.  The
$\pi_2$ values at 5.5, 5.6, and 5.7 are offset for clarity.}
\label{fig:spectrum} 
\end{figure}
\begin{table}
\caption{Summary of results for $N_t = 12$, $am_q = 0.008$:  
chiral condensate, Polyakov loop, fuzzy Polyakov loop and induced
baryon charge.
\label{tab:nt12m008a}
}
\begin{tabular}{ldddd}
$6/g^2$ & $\left\langle\bar\psi\psi\right\rangle$ 
  & $\mathop{\rm Re} P$ & $\mathop{\rm Re} F$ & $Q$ \\
\hline
5.65  & 0.0393(3)   & 0.0033(5)  & 0.00051(6)  & $-$0.6(2)  \\
5.70  & 0.0311(5)   & 0.0094(15) & 0.00155(11) & $-$0.05(7) \\
5.725 &	0.0290(2)   & 0.0118(17) & 0.00178(15) & $-$0.16(7) \\
5.75  &	0.0267(2)   & 0.0168(13) & 0.00255(12) & $-$0.03(3) \\
5.80  &	0.02424(10) & 0.0212(10) & 0.00359(10) & $-$0.03(2) \\
5.85  & 0.02262(7)  & 0.0254(12) &  --         &   --       
\end{tabular}
\end{table}
\begin{table}
\caption{Summary of results for $N_t = 12$, $am_q = 0.008$:
nonsinglet and singlet baryon susceptibility and disconnected and
connected chiral susceptibility.
\label{tab:nt12m008b}
}
\begin{tabular}{lddld}
$6/g^2$ & $\chi_{\rm ns}/T^2$ & $\chi_{\rm s}/T^2$ & 
 $\chi_{\rm disc}/T^2$ & $\chi_{\rm conn}/T^2$ \\
\hline
5.65  & 0.6(2)   & 0.4(5) &  54(10)   & 480(8)  \\
5.70  & 1.40(13) & 1.3(5) &  80(20)   & 466(12) \\
5.725 & 1.44(11) & 1.5(4) &  29(6)    & 469(9)  \\
5.75  & 1.52(10) & 1.6(3) &  25(7)    & 458(7)  \\
5.80  & 1.87(8)  & 1.8(4) &  6.2(1.1) & 435(6)  \\
5.85  &  --      &  --    &   --      &  --     
\end{tabular}
\end{table}
\begin{table}
\caption{Summary of results for $N_t = 12$, $am_q = 0.016$:
chiral condensate, Polyakov loop, fuzzy Polyakov loop and induced
baryon charge.
\label{tab:nt12m016a}
}
\begin{tabular}{ldddd}
$6/g^2$ & $\left\langle\bar\psi\psi\right\rangle$ 
   & $\mathop{\rm Re} P$ & $\mathop{\rm Re} F$ & $Q$ \\
\hline
5.65  & 0.0672(2)   & 0.0034(4)  & 0.00050(5)  & $-$0.2(2)  \\
5.70  & 0.0599(2)   & 0.0046(9)  & 0.00065(8)  & $-$0.4(2)  \\
5.75  &	0.0522(3)   & 0.0136(8)  & 0.00205(9)  & $-$0.12(4) \\
5.80  &	0.0491(3)   & 0.0164(12) & 0.00225(15) & $-$0.10(3) \\
5.85  & 0.04516(14) & 0.0236(13) & 0.00368(14) &  0.01(3) 
\end{tabular}
\end{table}
\begin{table}
\caption{Summary of results for $N_t = 12$, $am_q = 0.016$:
nonsinglet and singlet baryon susceptibility and disconnected and
connected chiral susceptibility.
\label{tab:nt12m016b}
}
\begin{tabular}{lddld}
$6/g^2$ & $\chi_{\rm ns}/T^2$ & $\chi_{\rm s}/T^2$ & 
 $\chi_{\rm disc}/T^2$ & $\chi_{\rm conn}/T^2$ \\
\hline
5.65  & 0.48(15) & 0.9(4) & 31(6)  & 434(6) \\
5.70  & 0.57(12) & 0.4(5) & 24(6)  & 427(5) \\
5.75  & 1.14(12) & 0.4(4) & 26(3)  & 427(7) \\
5.80  & 1.21(10) & 1.3(3) & 30(5)  & 404(4) \\
5.85  & 1.58(11) & 1.2(4) & 8(2)   & 382(5)  
\end{tabular}
\end{table}
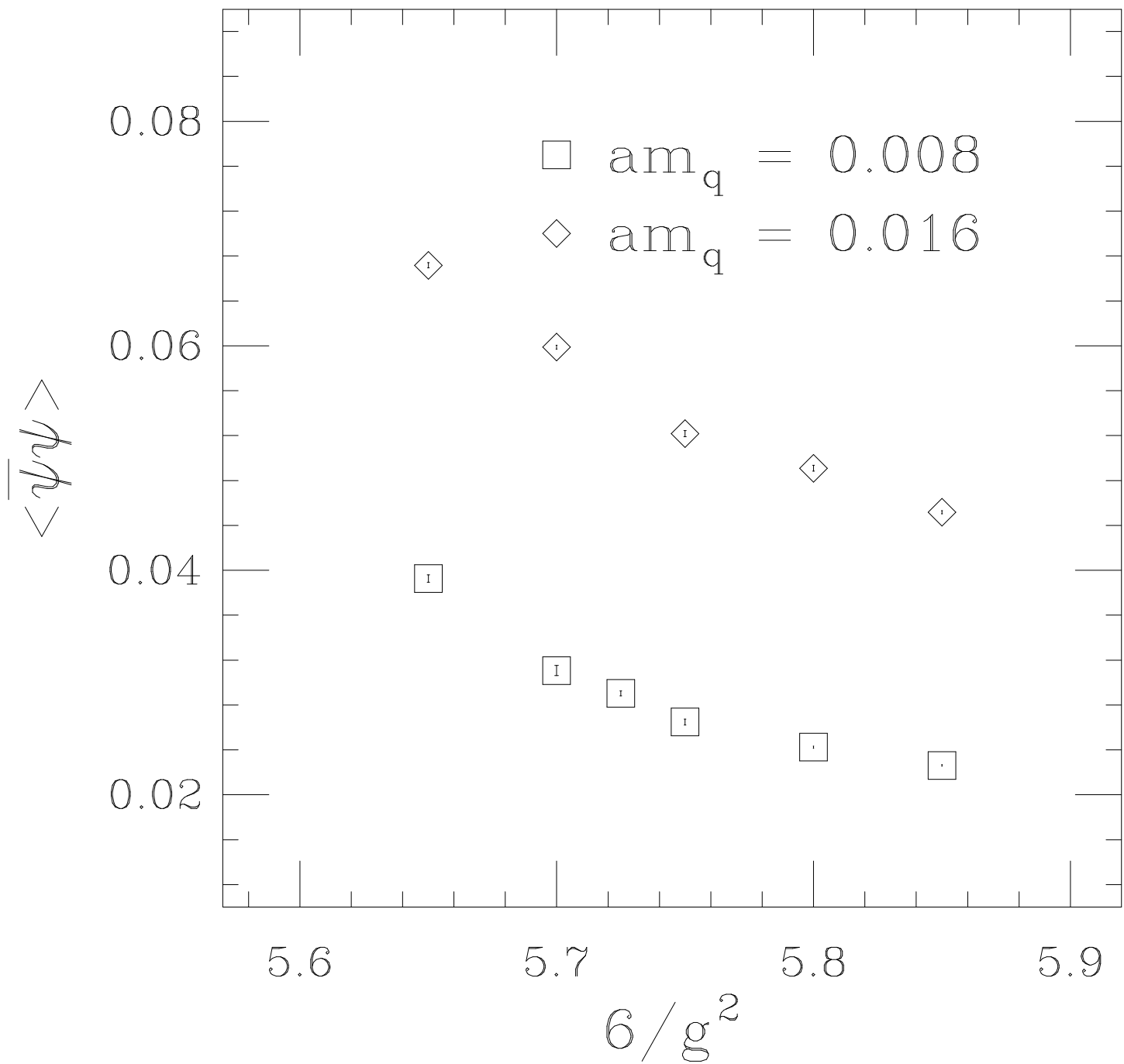
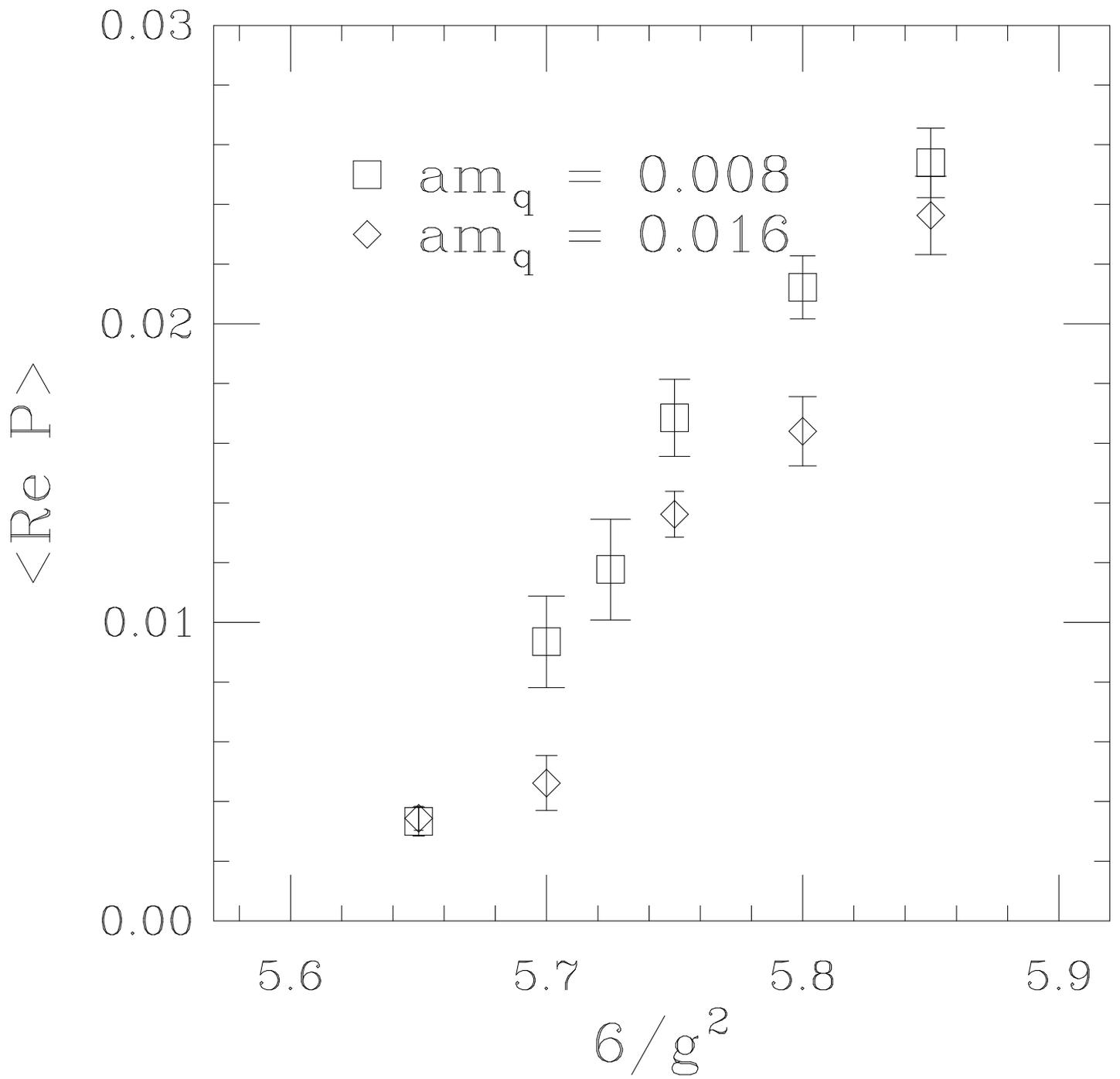
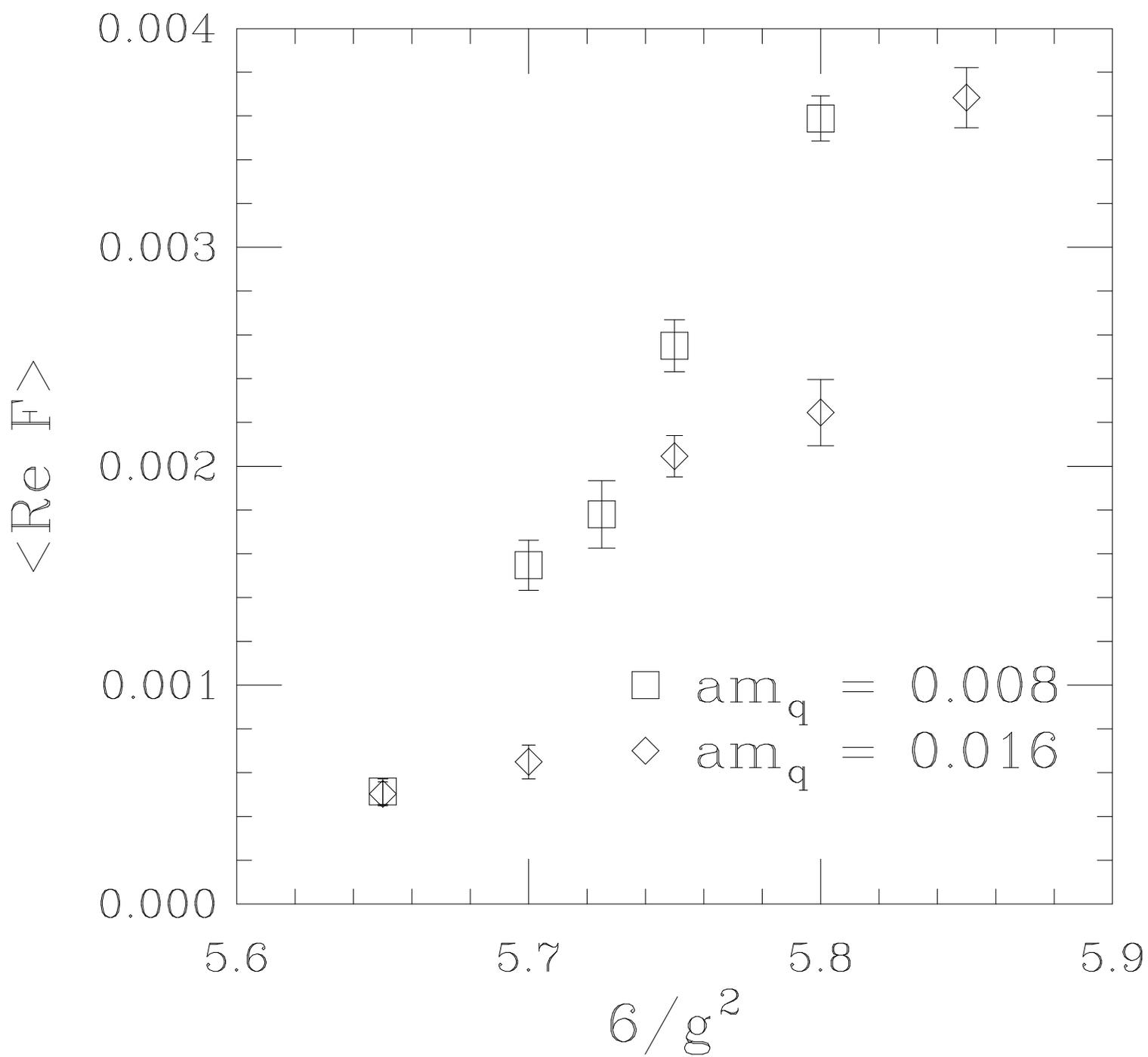
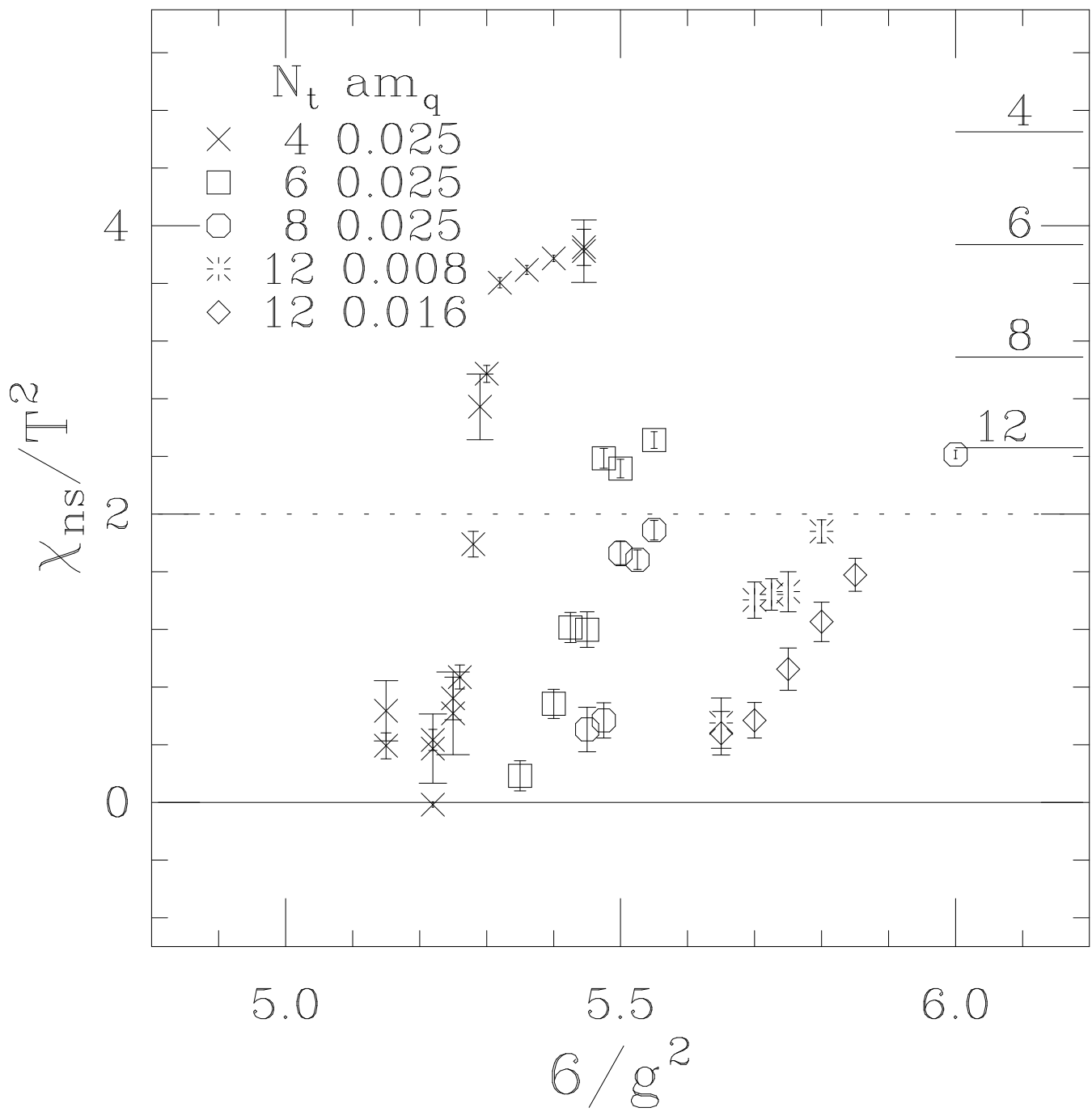
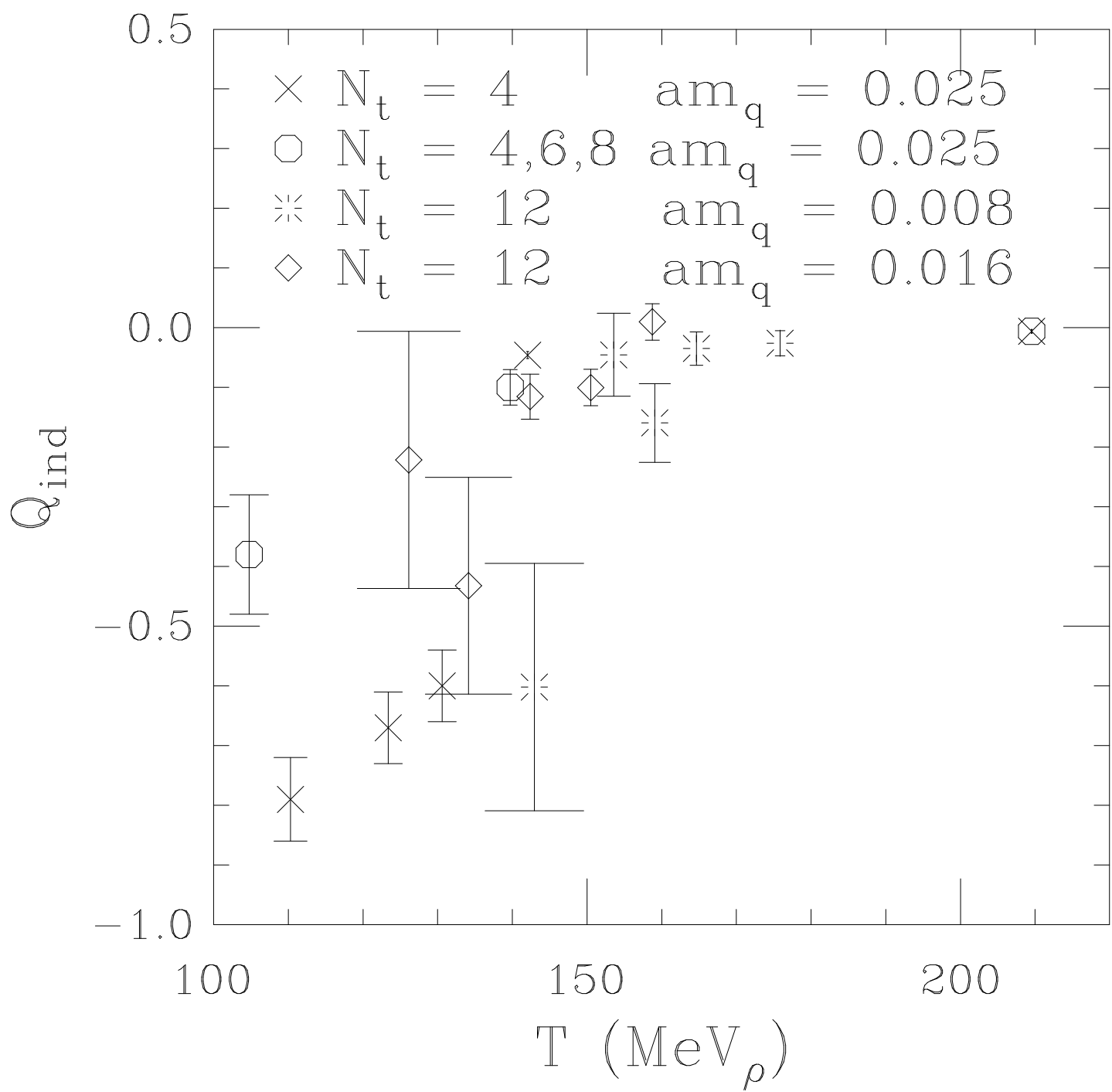
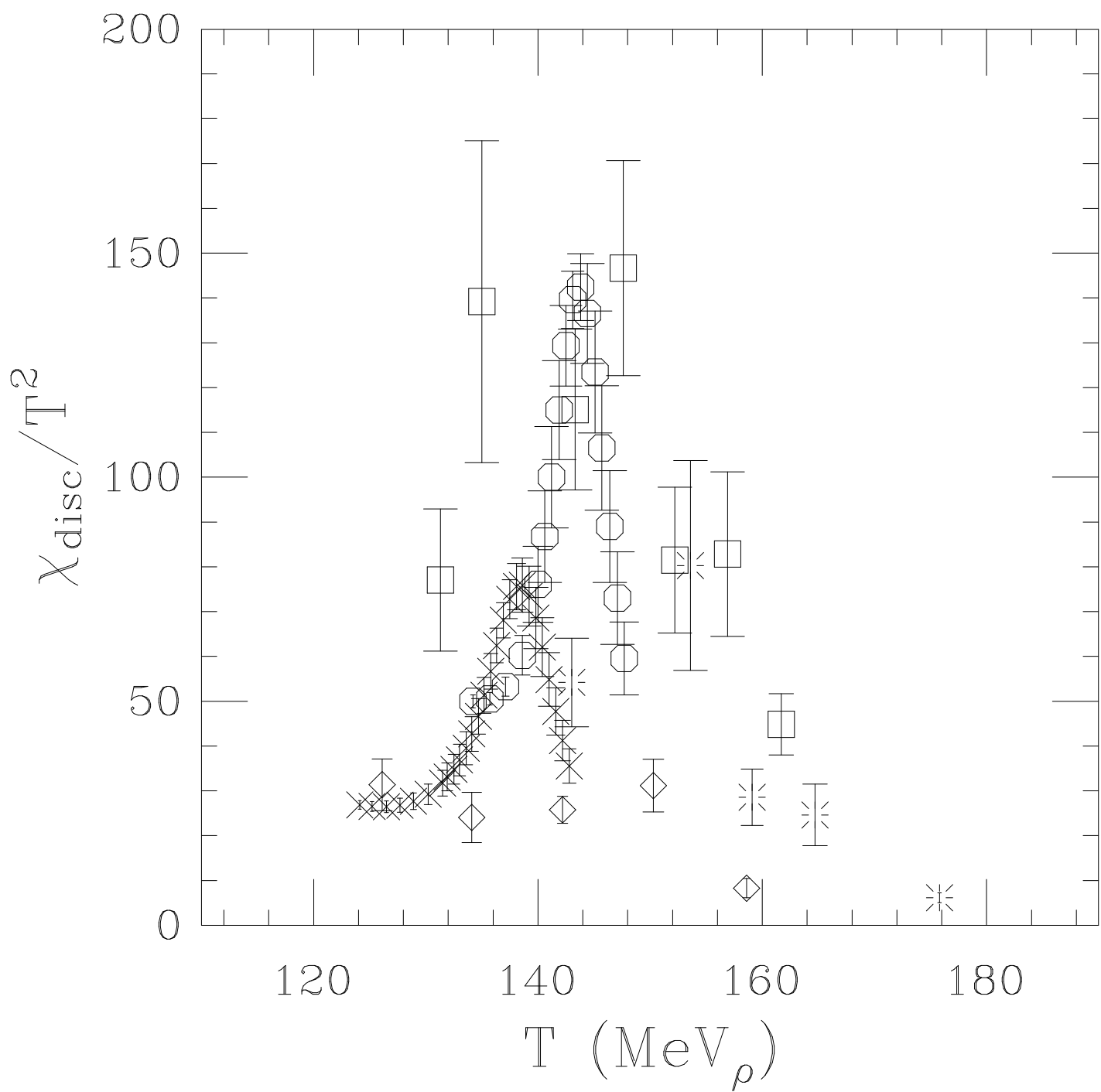
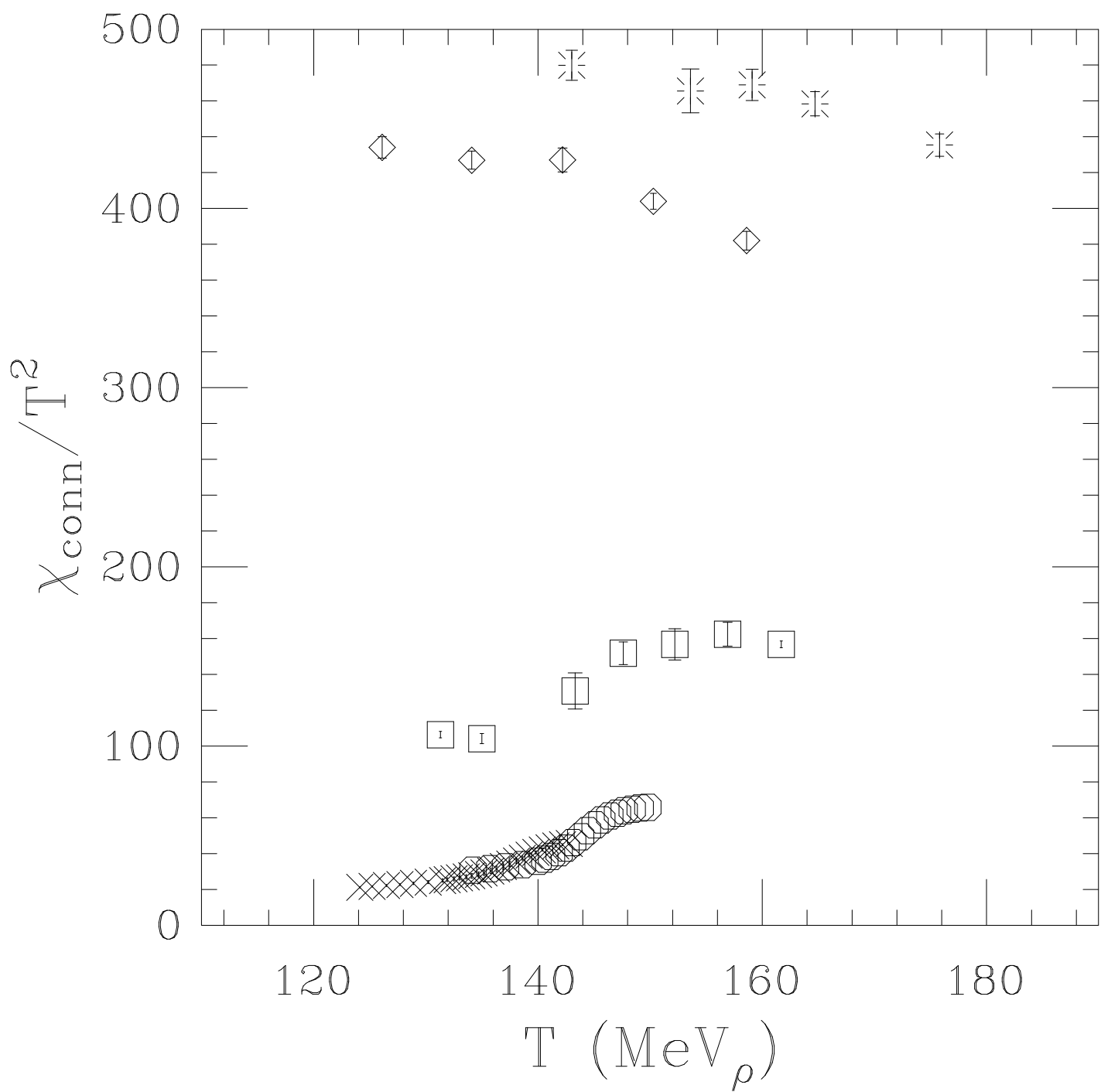
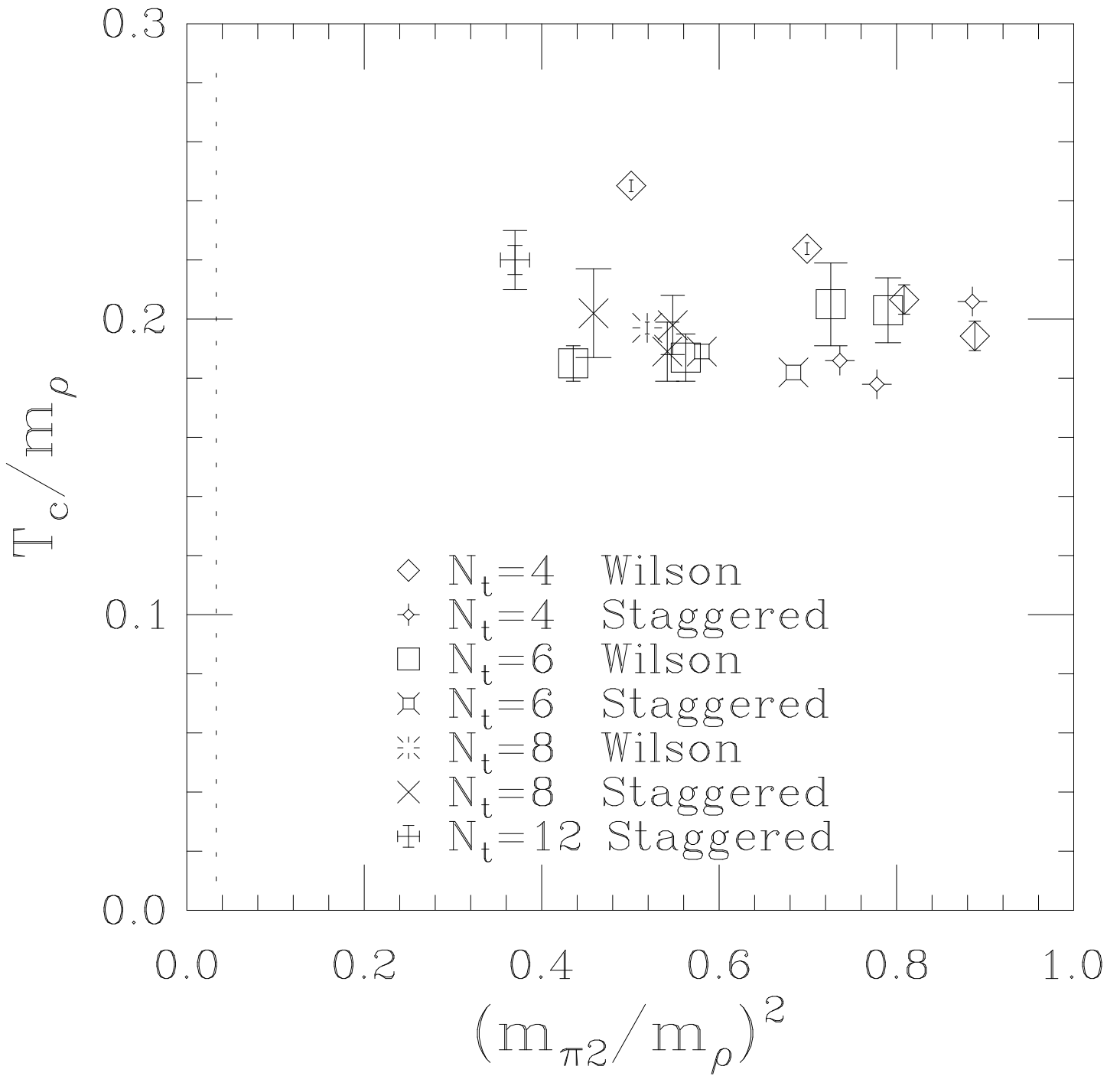
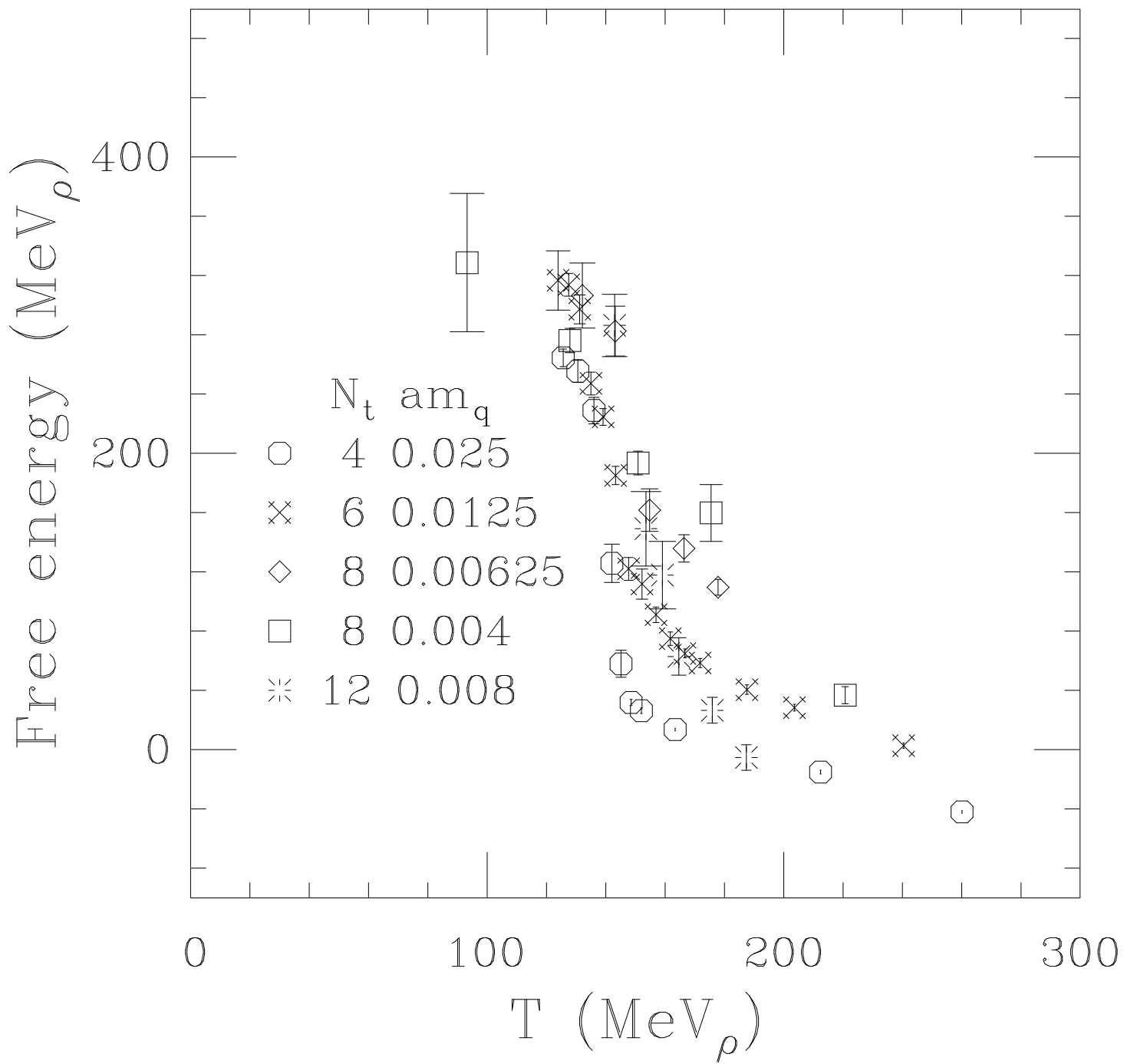
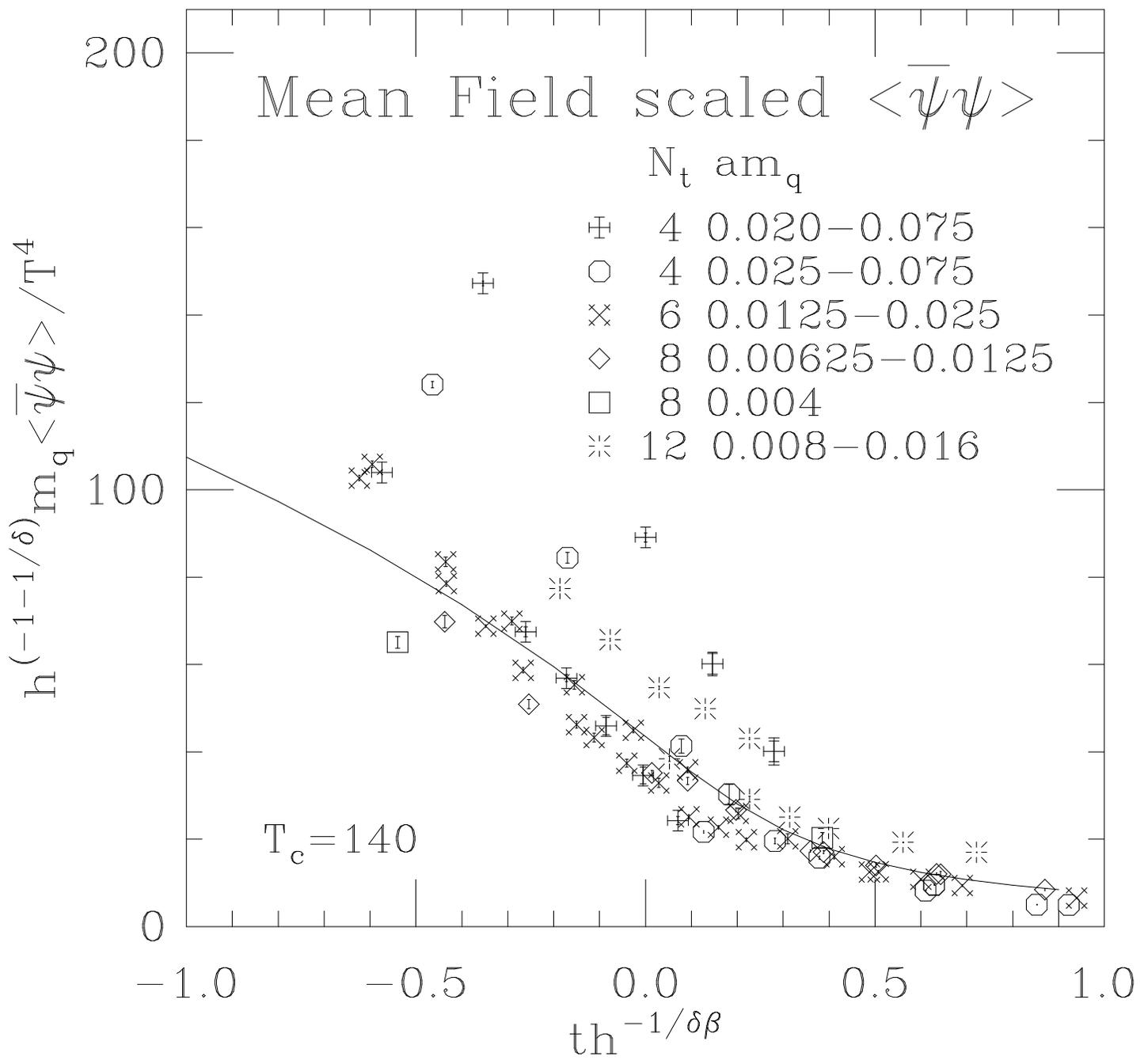
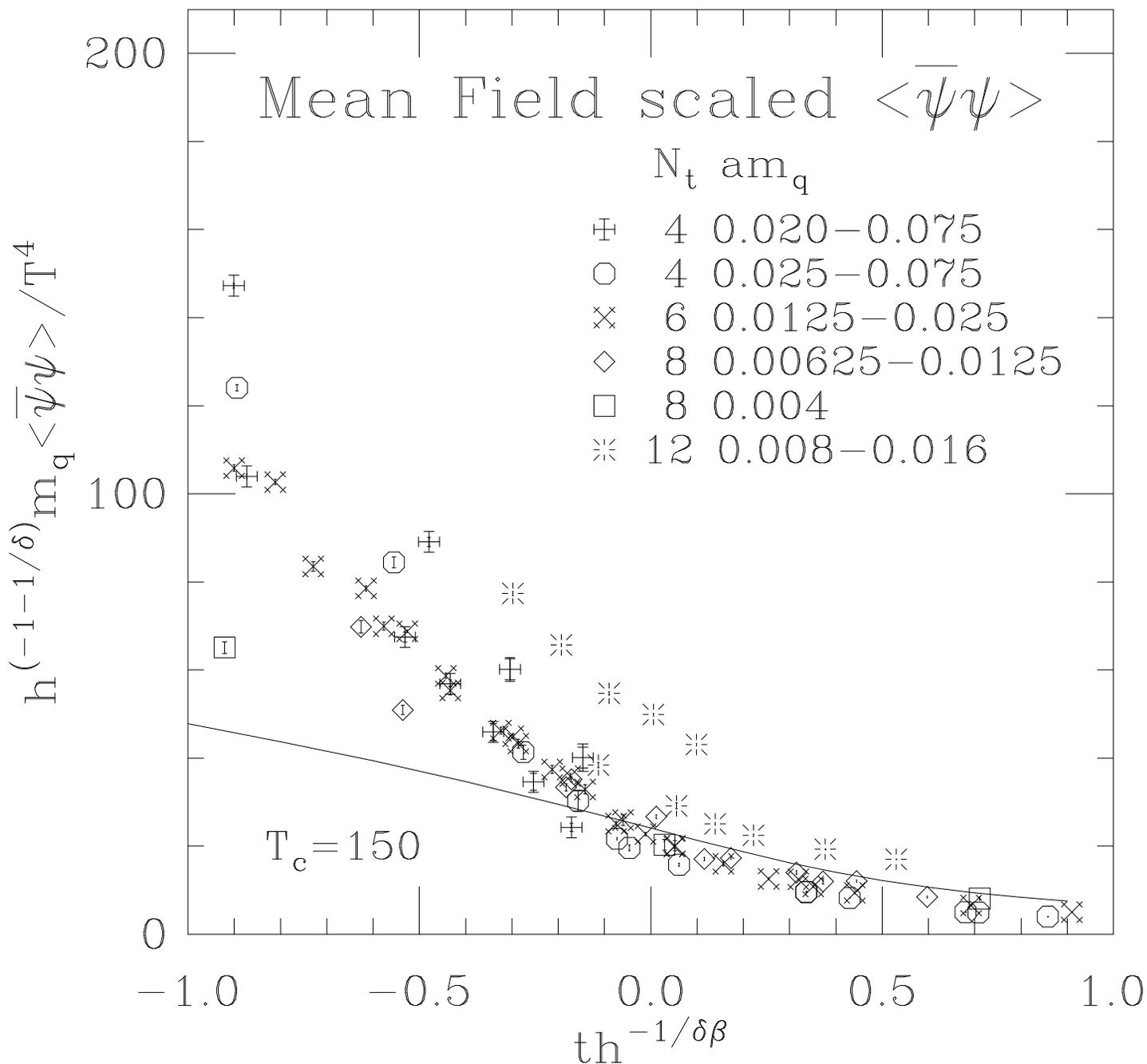
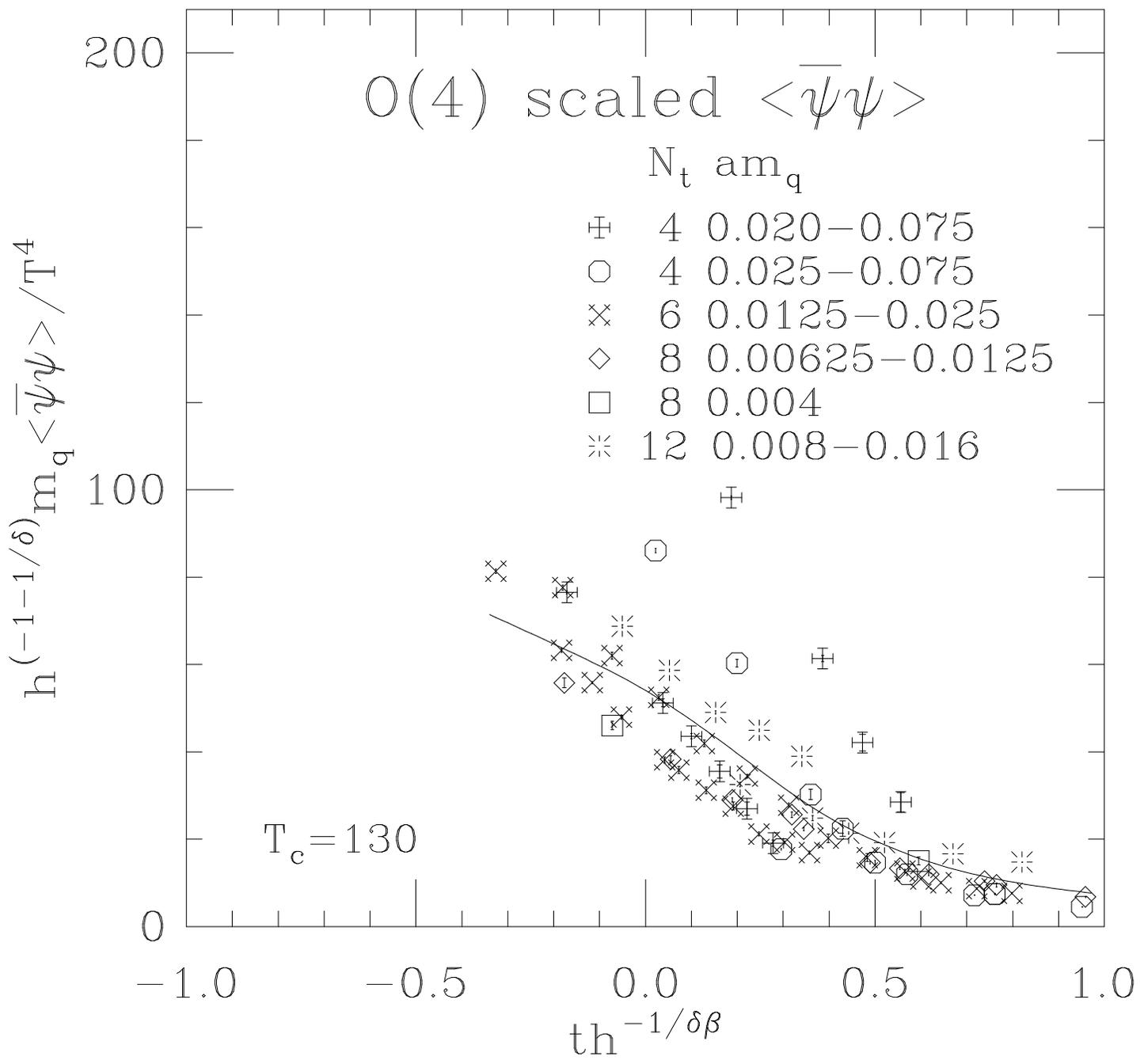
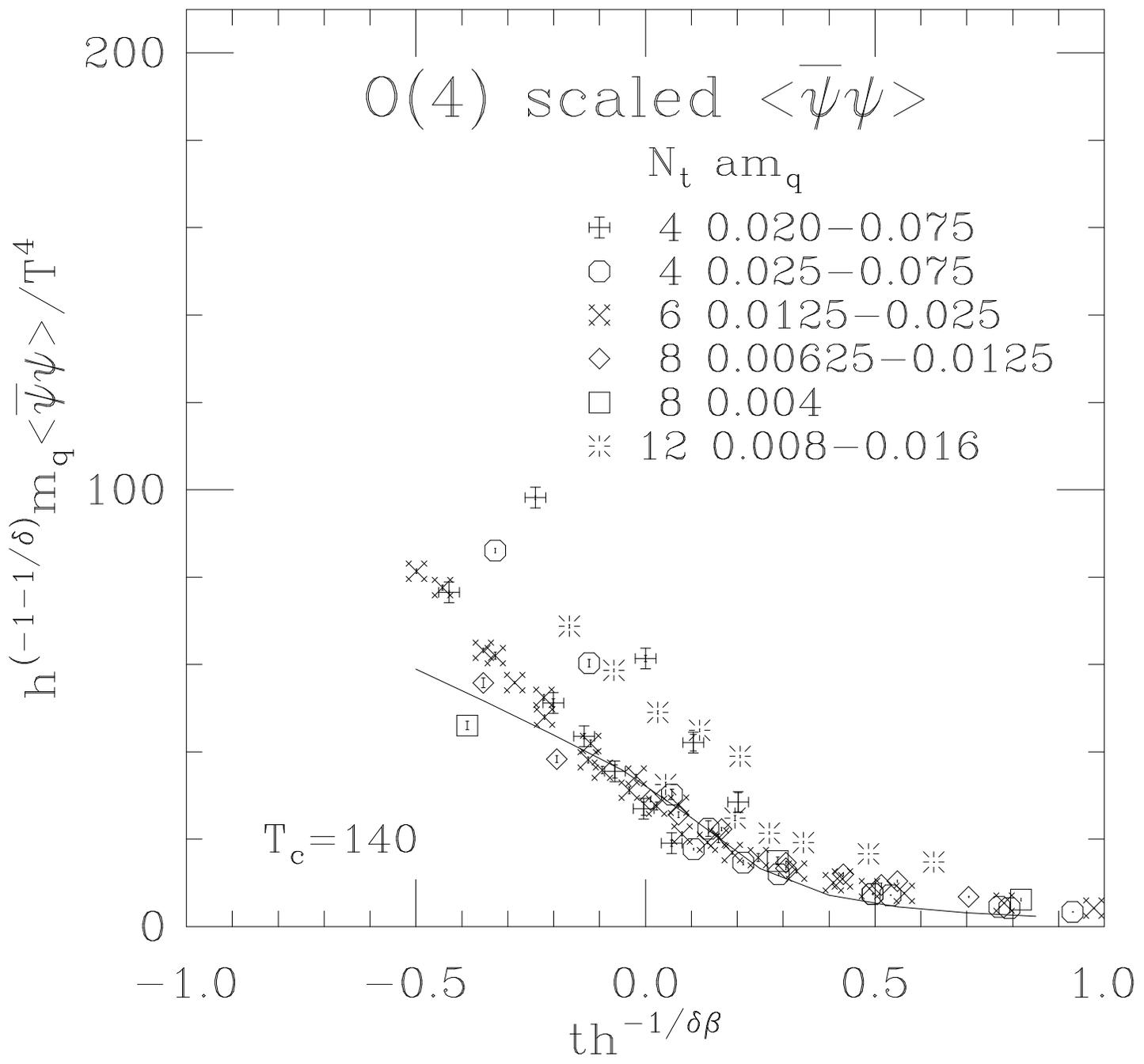
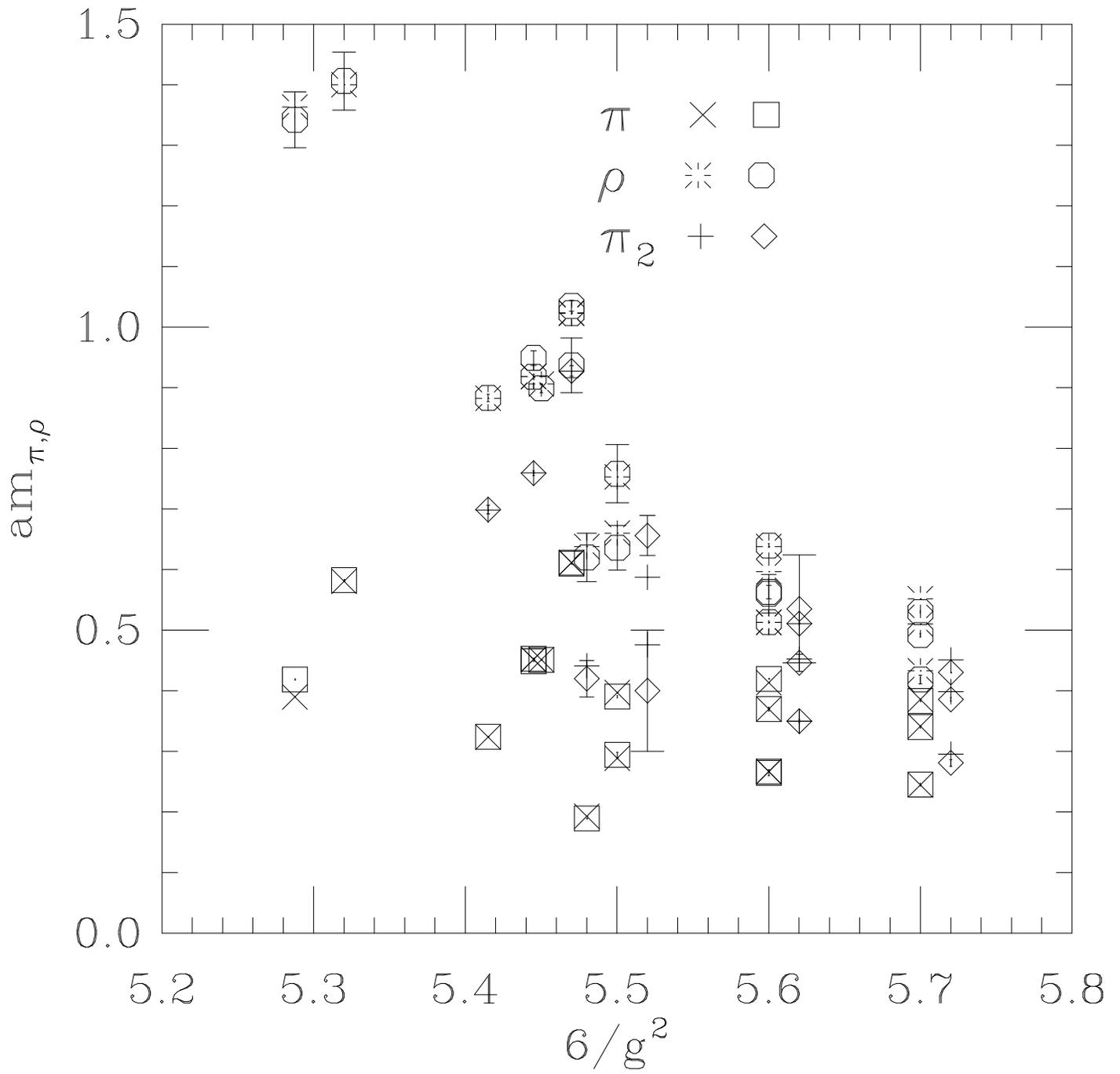
\end{document}